\documentclass[twoside,twocolumn,9pt]{article}
\usepackage{extsizes}
\usepackage[super,sort&compress,comma]{natbib} 
\usepackage[version=3]{mhchem}
\usepackage[left=1.5cm, right=1.5cm, top=1.785cm, bottom=2.0cm]{geometry}
\usepackage{balance}
\usepackage{mathptmx}
\usepackage{sectsty}
\usepackage{graphicx} 
\usepackage{lastpage}
\usepackage[format=plain,justification=justified,singlelinecheck=false,font={stretch=1.125,small,sf},labelfont=bf,labelsep=space]{caption}
\usepackage{float}
\usepackage{fancyhdr}
\usepackage{fnpos}
\usepackage[english]{babel}
\addto{\captionsenglish}{%
  
}
\usepackage{array}
\usepackage{droidsans}
\usepackage{charter}
\usepackage[T1]{fontenc}
\usepackage[usenames,dvipsnames]{xcolor}
\usepackage{setspace}
\usepackage[compact]{titlesec}
\usepackage{hyperref}
\usepackage{epstopdf}
\definecolor{cream}{RGB}{222,217,201}

\begin{document}

\pagestyle{fancy}
\thispagestyle{plain}
\fancypagestyle{plain}{
%%%HEADER%%%
\renewcommand{\headrulewidth}{0pt}
}
%%%END OF HEADER%%%

%%%PAGE SETUP - Please do not change any commands within this section%%%
\makeFNbottom
\makeatletter
\renewcommand\LARGE{\@setfontsize\LARGE{15pt}{17}}
\renewcommand\Large{\@setfontsize\Large{12pt}{14}}
\renewcommand\large{\@setfontsize\large{10pt}{12}}
\renewcommand\footnotesize{\@setfontsize\footnotesize{7pt}{10}}
\makeatother

\renewcommand{\thefootnote}{\fnsymbol{footnote}}
\renewcommand\footnoterule{\vspace*{1pt}% 
\color{cream}\hrule width 3.5in height 0.4pt \color{black}\vspace*{5pt}} 
\setcounter{secnumdepth}{5}

\makeatletter 
\renewcommand\@biblabel[1]{#1}            
\renewcommand\@makefntext[1]% 
{\noindent\makebox[0pt][r]{\@thefnmark\,}#1}
\makeatother 
\renewcommand{\figurename}{\small{Fig.}~}
\sectionfont{\sffamily\Large}
\subsectionfont{\normalsize}
\subsubsectionfont{\bf}
\setstretch{1.125} %In particular, please do not alter this line.
\setlength{\skip\footins}{0.8cm}
\setlength{\footnotesep}{0.25cm}
\setlength{\jot}{10pt}
\titlespacing*{\section}{0pt}{4pt}{4pt}
\titlespacing*{\subsection}{0pt}{15pt}{1pt}
%%%END OF PAGE SETUP%%%

%%%FOOTER%%%
\fancyfoot{}
\fancyfoot[LO,RE]{\vspace{-7.1pt}\includegraphics[height=9pt]{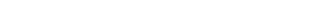}}
\fancyfoot[CO]{\vspace{-7.1pt}\hspace{13.2cm}\includegraphics{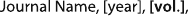}}
\fancyfoot[CE]{\vspace{-7.2pt}\hspace{-14.2cm}\includegraphics{head_foot/RF}}
\fancyfoot[RO]{\footnotesize{\sffamily{1--\pageref{LastPage} ~\textbar  \hspace{2pt}\thepage}}}
\fancyfoot[LE]{\footnotesize{\sffamily{\thepage~\textbar\hspace{3.45cm} 1--\pageref{LastPage}}}}
\fancyhead{}
\renewcommand{\headrulewidth}{0pt} 
\renewcommand{\footrulewidth}{0pt}
\setlength{\arrayrulewidth}{1pt}
\setlength{\columnsep}{6.5mm}
\setlength\bibsep{1pt}
%%%END OF FOOTER%%%

%%%FIGURE SETUP - please do not change any commands within this section%%%
\makeatletter 
\newlength{\figrulesep} 
\setlength{\figrulesep}{0.5\textfloatsep} 

\newcommand{\topfigrule}{\vspace*{-1pt}% 
\noindent{\color{cream}\rule[-\figrulesep]{\columnwidth}{1.5pt}} }

\newcommand{\botfigrule}{\vspace*{-2pt}% 
\noindent{\color{cream}\rule[\figrulesep]{\columnwidth}{1.5pt}} }

\newcommand{\dblfigrule}{\vspace*{-1pt}% 
\noindent{\color{cream}\rule[-\figrulesep]{\textwidth}{1.5pt}} }

\makeatother
%%%END OF FIGURE SETUP%%%

%%%TITLE, AUTHORS AND ABSTRACT%%%
\twocolumn[
  \begin{@twocolumnfalse}
{\includegraphics[height=30pt]{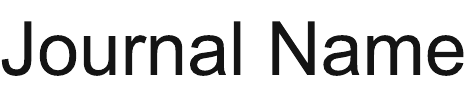}\hfill\raisebox{0pt}[0pt][0pt]{\includegraphics[height=55pt]{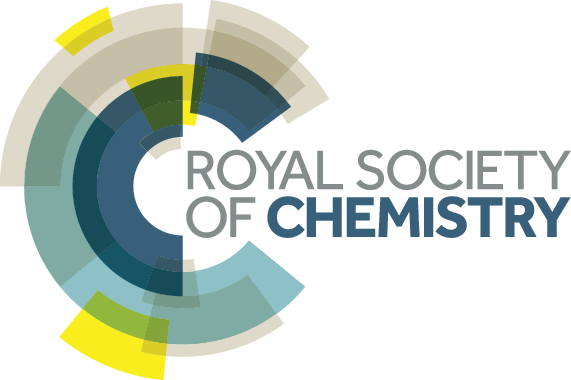}}\\[1ex]
\includegraphics[width=18.5cm]{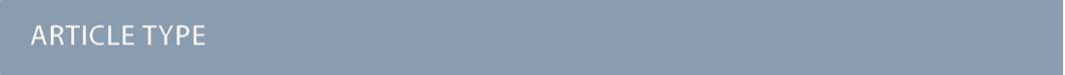}}\par
\vspace{1em}
\sffamily
\begin{tabular}{m{4.5cm} p{13.5cm} }

\includegraphics{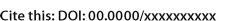} & \noindent\LARGE{\textbf{Irida-Graphene Phonon Thermal Transport via Non-equilibrium Molecular Dynamics Simulations}}
\\
\vspace{0.3cm} & \vspace{0.3cm}
\\

 & \noindent\large{Isaac M. Felix,\textit{$^{a}$}
Raphael M. Tromer,\textit{$^{b}$}
Leonardo D. Machado,\textit{$^{c}$}
Douglas S. Galvão,\textit{$^{d}$}
Luiz A. Ribeiro Jr,\textit{$^{e}$} and
Marcelo L. Pereira Jr\textit{$^{f, \ast}$}
 }
 \\

\includegraphics{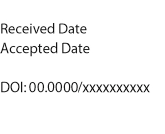} & \noindent\normalsize{Recently, a new 2D carbon allotrope called Irida-Graphene (Irida-G) was proposed. Irida-G consists of a flat sheet topologically arranged into 3-6-8 carbon rings exhibiting metallic and non-magnetic properties. In this study, we investigated the thermal transport properties of Irida-G using classical reactive molecular dynamics simulations. The findings indicate that Irida-G has an intrinsic thermal conductivity of approximately 215 W/mK at room temperature, significantly lower than that of pristine graphene. This decrease is due to characteristic phonon scattering within Irida-G's porous structure. Additionally, the phonon group velocities and vibrational density of states for Irida-G were analyzed, revealing reduced average phonon group velocities compared to graphene. The thermal conductivity of Irida-G is isotropic and shows significant size effects, transitioning from ballistic to diffusive heat transport regimes as the system length increases. These results suggest that while Irida-G has lower thermal conductivity than graphene, it still holds potential for specific thermal management applications, sharing characteristics with other two-dimensional materials.}

\end{tabular}

 \end{@twocolumnfalse} \vspace{0.6cm}

  ]
%%%END OF TITLE, AUTHORS AND ABSTRACT%%%

%%%FONT SETUP - please do not change any commands within this section
\renewcommand*\rmdefault{bch}\normalfont\upshape
\rmfamily
\section*{}
\vspace{-1cm}

%%%FOOTNOTES%%%

\footnotetext{\textit{$^{a}$~Department of Physics, Federal University of Pernambuco, Recife, Pernambuco, Brazil.}}
\footnotetext{\textit{$^{b}$~School of Engineering, MackGraphe, Mackenzie Presbyterian University, São Paulo, São Paulo, Brazil.}}
\footnotetext{\textit{$^{c}$~Department of Physics, Federal University of Rio Grande do Norte, Natal, Rio Grande do Norte, Brazil.}}
\footnotetext{\textit{$^{d}$~Department of Applied Physics and Center for Computational Engineering and Sciences, State University of Campinas, Campinas, São Paulo, Brazil.}}
\footnotetext{\textit{$^{e}$~University of Brasília, Institute of Physics, Brasília, Federal District, Brazil.}}
\footnotetext{\textit{$^{f}$~University of Bras\'{i}lia, College of Technology, Department of Electrical Engineering, Bras\'{i}lia, Federal District, Brazil.}}
\footnotetext{\textit{$^{\ast}$~Corresponding Author: marcelo.lopes@unb.br}}

\footnotetext{\dag~Supplementary Information available: [details of any supplementary information available should be included here]. See DOI: 00.0000/00000000.}

%%%END OF FOOTNOTES%%%

%%%MAIN TEXT%%%%
\section{Introduction}
The study of thermal transport properties in nanomaterials has attracted considerable interest from the scientific community due to their potential applications in nanoelectronics, thermoelectric devices, and thermal management systems \cite{radhakrishnan2024}. In particular, two-dimensional (2D) materials have been extensively investigated to understand their thermal conductivities better. These properties are crucial for efficient heat dissipation in miniaturized electronic components \cite{duan2023}. The unique characteristics of 2D materials stem from their reduced dimensionality, often leading to physical phenomena not observed in their three-dimensional counterparts \cite{huang2020}. These attributes make 2D systems up-and-coming for advanced technological applications \cite{liu2023}.

Graphene, a single-atom-thick material with a hexagonal arrangement of carbon atoms, has been extensively studied due to its exceptional thermal and electrical conductivity, mechanical strength, and flexibility \cite{salahdin2022,tiwari2020}. Since its experimental realization in 2004 \cite{novoselov2004}, researchers have extensively investigated graphene physical properties \cite{mbayachi2021} and other materials with similar topologies, such as transition metal dichalcogenides (TMDs) \cite{huang2020}, hexagonal boron nitride (h-BN) \cite{ogawa2023}, and silicene \cite{zhu2021}. These materials offer unique physical properties, broadening their potential applications across various technological domains.

For carbon-based systems, numerous new structures, often derived from graphene, are continually being proposed in the literature \cite{egbedina2022}. Carbon's versatility allows it to form new structures with diverse characteristics based on different topologies and hybridizations, leading to a wide range of applications \cite{speranza2021}. Although most of these structures are theoretical propositions, they hold promise for experimental realization using modern synthesis techniques \cite{kanungo2022}. Notable advances in this area include the synthesis of PHA-Graphene \cite{wang2015,fan2019}, $\gamma$-Graphyne \cite{baughman1987,hu2022}, biphenylene \cite{hudspeth2010,fan2021}, and fullerene networks \cite{berber2004,hou2022}. All these structures were proposed before their synthesis.

Despite significant progress in nanomaterials, the quest to discover and characterize new 2D systems with novel physical and chemical properties continues \cite{paras2022}. Our group recently proposed the Irida-Graphene (Irida-G), a new carbon-based nanomaterial \cite{junior2023}, which has since been the focus of several studies \cite{zhang2024,tan2024stable,majidi2024irida,li2023,majidi2024twin}. Initially, we described Irida-G as having a flat structure composed of 3, 6, and 8-membered rings, with no buckling and a cohesive energy of -7.0 eV/atom. It exhibits structural and thermal stability at temperatures up to 1000 K. Irida-G is metallic and non-magnetic, with Young's modulus of approximately 400 GPa. It can withstand temperatures up to 4000 K, its sublimation point. Optically, Irida-G is active in the infrared and ultraviolet regions and has low reflectivity indices.

As mentioned above, other researchers have carried out further studies on Irida-G. For instance, Zhang and Guo investigated the effect of lithium decoration on Irida-G for potential hydrogen storage applications \cite{zhang2024}. Their findings revealed that while H$_2$ adsorption on pristine Irida-G is weak, it is significantly enhanced when the material is decorated with lithium. Lithium atoms strongly interact with the octagonal carbon rings, preventing atom aggregation. Li-decorated Irida-G can adsorb up to 16 H$_2$ molecules per unit cell on one face and up to 24 on both faces, with suitable adsorption energies for reversible hydrogen storage. The storage density reaches 7.06 wt\%, surpassing the 6.5 wt\% standards set by the U.S. Department of Energy (DOE), indicating its promise for mobile hydrogen storage applications \cite{zhang2024}.

In another contribution, Tan and collaborators studied Irida-G decorated with titanium \cite{tan2024}. Their results showed that titanium atoms strongly bind to the Irida-G hexagonal rings, unlike lithium, which prefers the octagonal ones. A single titanium atom in the primitive unit cell can bind to 5 H$_2$ molecules, with an average adsorption energy of 0.41 eV/H$_2$, achieving a gravimetric density of 7.7 wt\%, again exceeding the DOE standard. The system's stability is maintained through Kubas-type interactions and a diffusion energy barrier of 5.0 eV, which prevents Ti-Ti aggregation \cite{tan2024}. Their ab initio molecular dynamics simulations demonstrated that the system remains stable at 600 K, above the desorption temperature of 524 K, indicating its stability during hydrogen recharge and discharge cycles \cite{tan2024}.

Inspired by the synthesis of $\gamma$-graphyne \cite{hu2022}, Majidi proposed inserting acetylenes into all the bonds of the original Irida-G, creating a new material called Irida-Graphyne (IGY) \cite{majidi2024irida}. The results demonstrated that IGY possesses confirmed thermal and dynamic stabilities, as evidenced by cohesive energy value, ab initio molecular dynamics simulations, and phonon dispersion data. Like the sp$^2$-hybridized Irida-G, this graphyne-like system is metallic. It exhibits strong anisotropic photoresponse under light irradiation with parallel and perpendicular polarizations. IGY showed high dielectric constants and optical absorption, suggesting its potential for use in energy storage systems, with broad absorption from the infrared to ultraviolet regions. Reflection and transmission data indicated that Irida-G and IGY are transparent, especially at high energies, making them suitable for optoelectronic devices. In another study, Majidi and Ayesh proposed Twin Irida-G, which was analyzed for its structural and thermal stability and electronic and optical properties \cite{majidi2024twin}. Although the stiffness in the Twin Irida-G plane is lower than that of graphene, it retains the same potential applications as single-layer Irida-G.

Li and colleagues investigated the mechanical properties of double-layer and single-layer Irida-G structures and the impact of nanocracks \cite{li2023}. Their findings revealed that the fracture stress of the double-layer Irida-G structure is higher than that of the single-layer structure. Analysis of cracks in both materials indicated that cracks perpendicular to the direction of the applied strain reduce fracture stress as the crack length increases. Additionally, the authors noted that larger angles relative to the strain direction further decrease fracture stress in the presence of nanocracks.

Understanding the thermal properties of 2D systems is crucial for advancing nanotechnologies, particularly because efficient heat management becomes increasingly important as electronic devices shrink in size. A detailed knowledge of a material's physical and chemical characteristics can also facilitate its synthesis and identify promising applications.

In this study, we carried out classical reactive molecular dynamics (MD) simulations to explore the thermal transport properties of Irida-G. Our investigation revealed that the intrinsic thermal conductivity of Irida-G at room temperature is about 215 W/mK, much lower than pristine graphene's value of around 1200 W/mK. This decrease is primarily attributed to enhanced phonon scattering from its porous structure. Furthermore, we computed the vibrational density of states and phonon group velocities, finding that Irida-G's average phonon group velocity is about 410 m/s compared to graphene's 780 m/s, providing additional evidence for the observed decrease in thermal conductivity. It is worth mentioning that the methodology employed here, as discussed below, is well-established and has been used in numerous other studies \cite{Felix2018,mortazavi2014,mortazavi2013,mortazavi2016}.

\section{Methodology}

We used the well-known LAMMPS software for the MD simulations in this study \cite{Thompson2022}. Interatomic forces were described using the second-generation REBO potential \cite{brenner2002}, extensively validated for various 2D carbon systems \cite{ong2011,mortazavi2014,mu2014,barbarino2015,wei2022}. To validate its accuracy for Irida-G's atomic bonding, we computed phonon dispersions using GULP \cite{gale1997,gale2003}. We also tested the original Tersoff potential \cite{Tersoff1988} and Lindsay's modified version \cite{Lindsay2010}, but neither produced good Irida-G's results.

In all MD simulations, the equations of motion were integrated using a timestep of 0.25 fs. The systems were initially thermalized using a Nosé-Hoover thermostat at 300 K for 100 ps. Periodic boundary conditions were applied along the xy-plane, while free boundary conditions were imposed along the z-direction. Each system underwent relaxation at finite temperatures until zero stress was achieved along the periodic directions; stress along the z-direction automatically reached zero. After equilibration, the thermostat was deactivated, allowing the systems to evolve under microcanonical conditions.

\begin{figure*}[t]
\begin{center}
\includegraphics[width=0.7\linewidth]{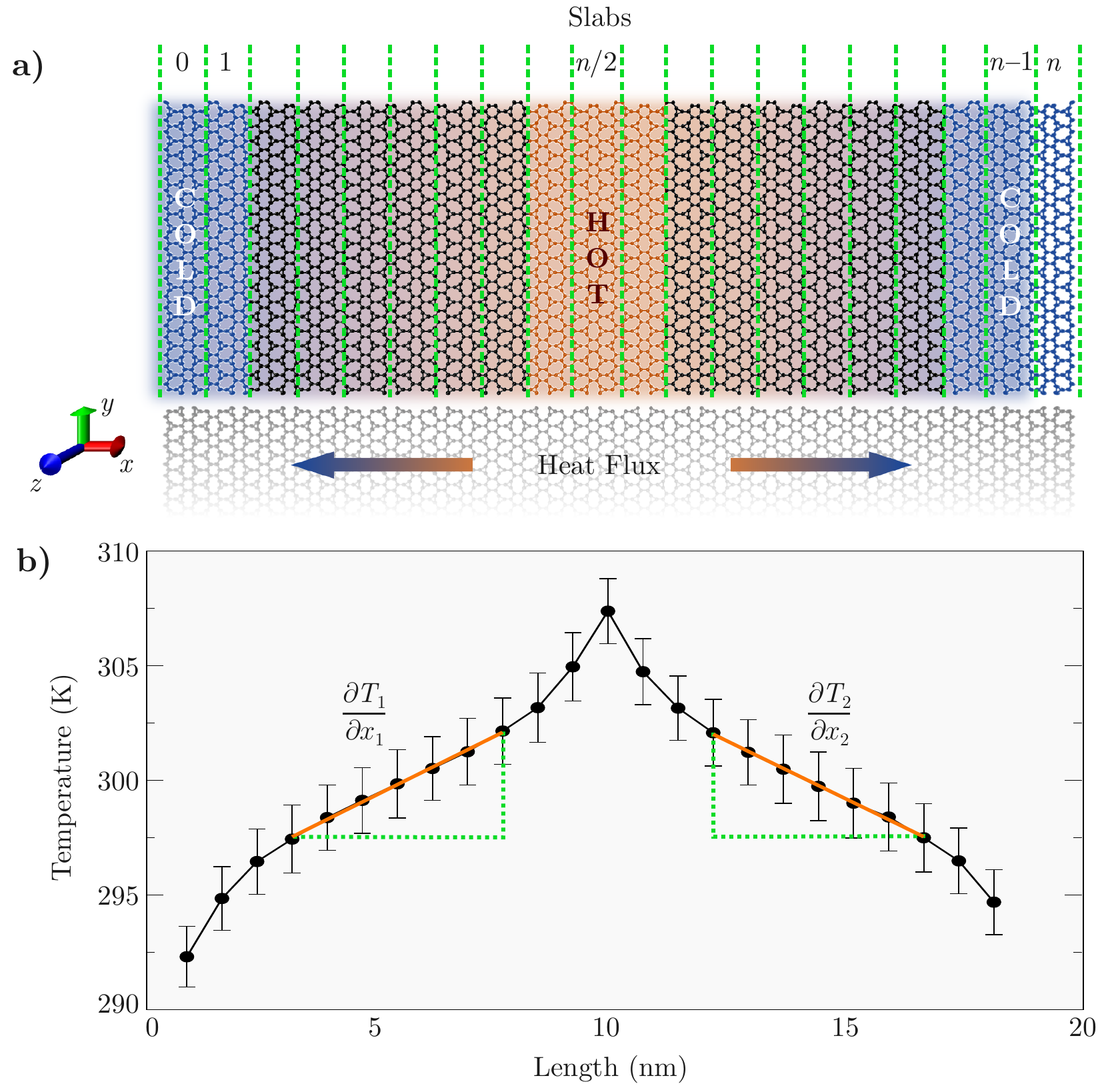}
\caption{
Set up for the RNEMD method: (a) The heat flux is applied along the Irida-G x-direction by exchanging kinetic energy between slower-moving particles in the cold region and faster-moving particles in the hot region. (b) A typical temperature profile is shown for an Irida-G sheet measuring 20 nm × 20 nm under steady-state conditions.
}
\label{fig:rnemd-method}
\end{center}
\end{figure*}

The reverse nonequilibrium molecular dynamics (RNEMD) method \cite{Muller1997} was employed to induce a heat flux within the system. This methodology, which we have previously used on thermal transport in various 2D systems \cite{Felix2018,Tromer2020,Tromer2020pccp,pereira2022}, partitions the system into $n$ slabs along its length, with two of these slabs designated as ``hot'' and ``cold'' regions, as illustrated in Figure \ref{fig:rnemd-method}(a). In our setup, the first slab was designated as the cold region, while the middle slab served as the hot region. Due to periodic boundary conditions, slabs $0$ and $n$ were identical. Each slab contained an average of $200-300$ atoms.

The heat flux is induced by exchanging kinetic energy between slow-moving particles in the hot region and fast-moving particles in the cold region, thereby establishing a temperature gradient.
%Kinetic energy exchanges occurred every 1000 simulation steps throughout $20$ ns (equivalent to $80\times10^6$ steps).
The heat flux is computed based on the exchanged kinetic energy between the designated layers, as per \cite{Muller1997}:
\begin{equation}
J(t) =\frac{1}{2 \Delta t A} \; {\sum_{\mathrm{swaps}} \left[ K_{\mathrm{source}} - K_{\mathrm{sink}} \right] },
\end{equation}
\noindent where $\Delta t$ represents the time interval from the start of swaps, $K_\mathrm{source}$ and $K_\mathrm{sink}$ denote the kinetic energies of atoms in the heat source and sink, respectively. The summation encompasses all swaps. Here, $A$ denotes the cross-sectional area of the sheet, calculated as the product of its width and thickness. All simulated systems shared a standard width of $20$ nm along the $y$-direction, with a thickness assumed to be $0.33$ nm along the $z$-direction. The factor of $2$ accommodates the distribution of heat flux between both sides of the heat source. In this investigation, energy swaps were executed every $500$ timesteps, covering a total simulation duration of $80$ million timesteps, equivalent to $20$ ns.

The simulations encompassed systems of considerable size, ranging from 13,824 to 1,105,920 atoms. These systems varied in nominal lengths from 10 nm to 800 nm, covering different sizes. This broad scope ensured the robustness and reliability of our simulations by providing statistically significant atom representation and comprehensive coverage of structural dynamics.

After the system reaches its stationary state, marked by a constant heat flux, we estimate the temperature within each slab by averaging the kinetic energy of particles using the energy equipartition theorem \cite{Muller1997}:
\begin{equation}
T_i = \frac{1}{3 N_i k_B}\sum\limits_{j=1}^{N_i}\frac{p_j^2}{m_j},
\end{equation}
\noindent where $T_i$ represents the temperature of the $i$-th slab, $N_i$ is the number of atoms in it, $k_B$ is Boltzmann's constant, $m_j$ is the mass of atom $j$, and $p_j$ refers for its linear momentum. The temperature gradient of the system is then obtained from the average temperature in each slab.

Once the heat flux and temperature gradient are steady, the thermal conductivity for a sample of length $L$ can be directly obtained from Fourier's law:
\begin{equation}
\kappa(L) =  \frac{\langle J \rangle}{ \left \langle {\partial T}/{\partial x}\right\rangle},
\label{eq:fourier}
\end{equation}
\noindent where ${\partial T}/{\partial x}$ is the arithmetic mean of the temperature gradient along $x$-direction considering both sides of heat flux, as shown in Figure \ref{fig:rnemd-method} (b). The brackets $\langle \ \rangle$ represent the time average of the quantities.

\section{Results}

\subsection{Phonon Dispersion}

\begin{figure}[t]
\begin{center}
\includegraphics[width=\linewidth]{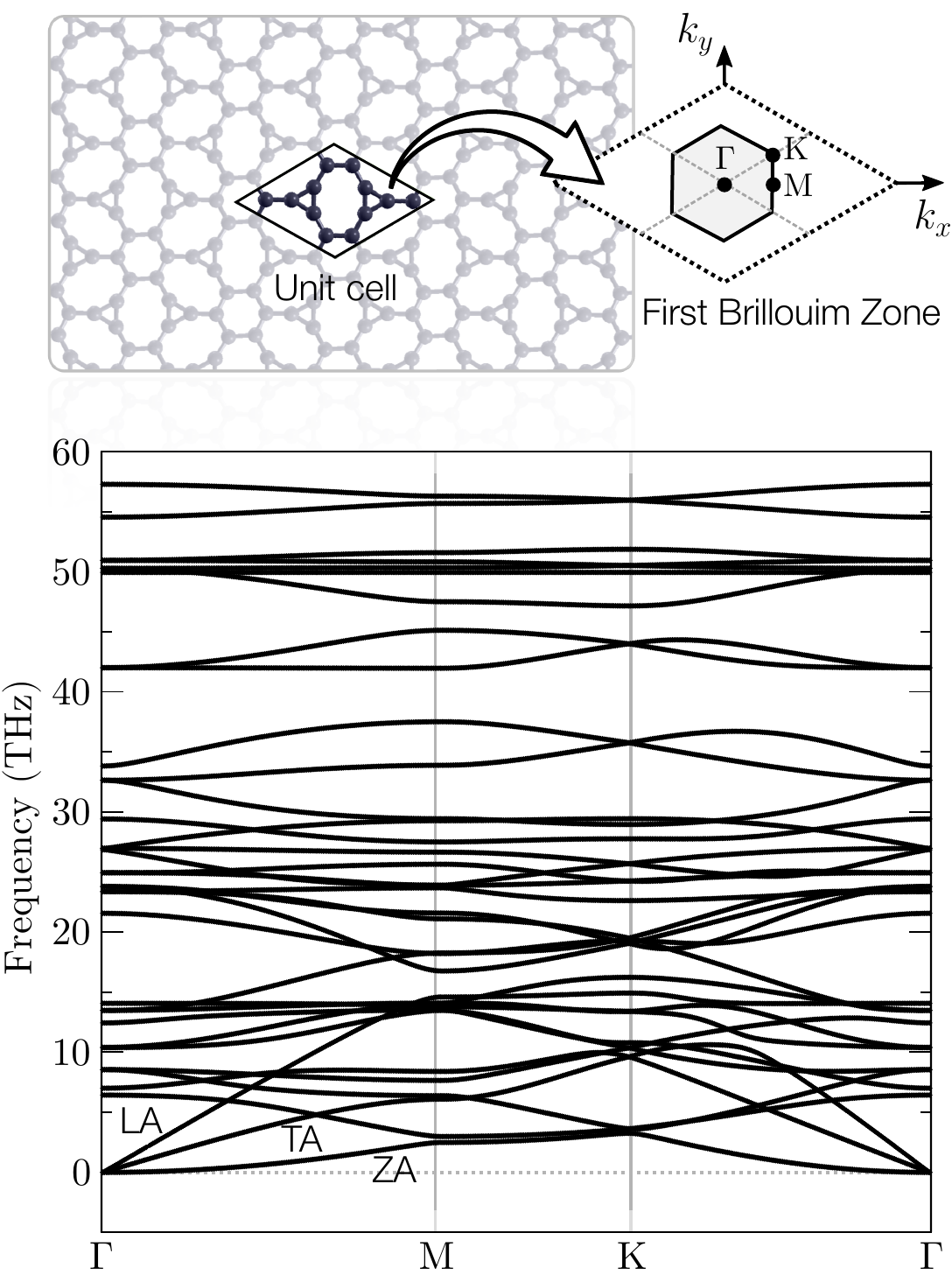}
\caption{Phonon dispersion for Irida-G was modeled using the second generation of REBO potential. The absence of negative (imaginary) frequencies indicates the stability of the structure with the chosen potential parameters. The inset shows the high symmetry points in reciprocal space coordinates: $\Gamma = (0,0,0)$, $\text{M}=(1/2,0,0)$, $\text{K}=(1/3,1/3,0)$ and $\Gamma = (0,0,0)$.}
\label{fig:irida-disper}
\end{center}
\end{figure}

To assess the accuracy of the second-generation REBO potential in describing the atomic bonding structure of Irida-G, we computed its phonon dispersions by diagonalizing the dynamical matrix using the lattice dynamics software GULP \cite{gale1997,gale2003}. Figure \ref{fig:irida-disper} illustrates the phonon dispersions along high symmetry points of the Brillouin Zone, obtained from the unit cell containing 12 carbon atoms. The absence of phonon modes with negative (imaginary) frequencies in the dispersion indicates the stability of Irida-G's crystal structure when modeled with the chosen potential.

The second-generation REBO potential typically accurately describes acoustic phonons but not so well the optical phonons. However, since acoustic phonons are the primary heat carriers in 2D systems \cite{lindsay2010_flexural,mu2014,barbarino2015,xu2015,pereira2016,taheri2021}, this force field can reliably simulate thermal transport in our study. Nevertheless, the overall thermal conductivity value obtained with this potential is expected to be lower than those obtained using ab initio methods that employ the Boltzmann transport equation. In Figure \ref{fig:irida-disper}, among the three acoustic phonon modes, both in-plane modes (longitudinal (LA) and transverse (TA)) exhibit linear dispersions. In contrast, the out-of-plane mode (flexural (ZA)) displays a quadratic dispersion around the $\Gamma$ point. Conversely, optical branches demonstrate relatively flat dispersions, indicating that phonons in these branches have small group velocities and do not contribute significantly to thermal transport.

\subsection{Thermal Conductivity Dependence with Sample Length}

\begin{figure}[t]
\centering
\includegraphics[width=\linewidth]{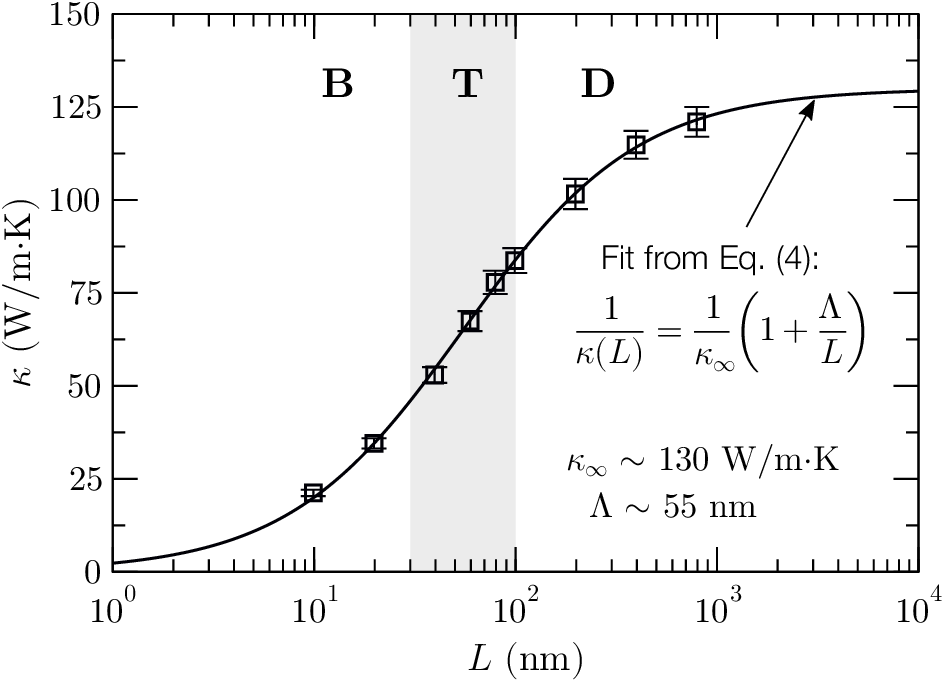}
\caption{Thermal conductivity of Irida-G as a function of sample length. Data points from RNEMD simulations and curve from Eq. (\ref{eq:kappa-length}).}
\label{fig:kappa}
\end{figure}

Figure \ref{fig:kappa} illustrates that the lattice thermal conductivity varies depending on the system length. Through error propagation from the average heat flux and temperature gradient, uncertainty analysis consistently yielded uncertainties below 5\% for all data points. Strong size effects emerge due to phonon confinement, which means the heat reservoirs delimit accessible paths within the region. This behavior is consistent with the theoretical framework proposed by Schelling and co-authors \cite{schelling2002}:
\begin{equation}
\label{eq:kappa-length}
    \frac{1}{\kappa(L)}=\frac{1}{\kappa_\infty}\left(1+\frac{\Lambda}{L}\right).
\end{equation}
\noindent Here, $\kappa_\infty$ represents the material's intrinsic (length-independent) conductivity, and $\Lambda$ denotes the effective mean free path of the heat carriers. Therefore, by fitting the above expression to the simulation data obtained for systems of increasing length, we can estimate both $\kappa_\infty$ and $\Lambda$ for Irida-G.

In the investigation performed here, the intrinsic lattice thermal conductivity at room temperature is predicted to be approximately 130 W/mK. At the same time, the effective accessible phonon mean free paths are around 55 nm. The fitting was conducted without the two largest systems, which were obtained later and presented more significant uncertainties but aligned with the fitted lines within the error bars. In fact, including the two largest systems in the fitting of Eq. (\ref{eq:kappa-length}) would result in a change of less than 5\% in the intrinsic lattice thermal conductivity $\kappa_\infty$. This agreement illustrates the remarkable predictive power of Eq. (\ref{eq:kappa-length}), enabling the estimation of intrinsic lattice thermal conductivity from simulations with relatively small systems \cite{Felix2018,Felix2020,Tromer2020,felix2022}.

We present here only the results for the heat flux along the $x$-direction (see Figure \ref{fig:rnemd-method}). However, all calculations were also performed for the $y$-direction, indicating that Irida-G is practically isotropic concerning the material’s intrinsic conductivity (134 W/mK) and effective mean free path of the heat carriers (58 nm).

When analyzing the behavior of $\kappa(L)$, we observe three distinct heat transport regimes, as shown in Figure \ref{fig:kappa}. Initially, in the ballistic regime (region B), characterized by small $L$ (up to approximately 30 nm), $\kappa(L)$ scales linearly with $L$. Here, the phonon mean free path exceeds the system length, constraining heat transfer. Conversely, for $L > 100$ nm, the diffusive regime emerges (region D), showcasing a weak dependency of thermal conductivity on system length. This phase's phonon mean free path falls short of the system length. Between these two regimes lies the ballistic-diffusive transition regime (region T), where the system length approaches the phonon mean free path, leading to a decreased dependence on thermal conductivity on sample length. Thus, phonons with intrinsic mean free paths longer than the sample's characteristic length exhibit ballistic transport. In contrast, those with shorter intrinsic mean free paths facilitate diffusive heat transfer. This behavior is consistent with the findings of Bae et al., who observed a ballistic to diffusive crossover of heat conductance in graphene ribbons of varying sizes \cite{bae2013}.

The intrinsic thermal conductivity of Irida-G is at least one order of magnitude smaller than the approximately 1200 W/mK obtained for pristine graphene using the same empirical potential (see supplementary material). This significant difference in $\kappa_\infty$ between Irida-G and graphene can be qualitatively attributed to the heightened phonon scattering induced by the structural porosity, characterized by triangles and octagons, within Irida-G. This trend increases the scattering of heat carriers, reducing the thermal conductivity of this material. Molecular dynamics simulations have similarly predicted a decreased thermal conductivity relative to graphene for both penta-graphene \cite{xu2015} and phagraphene \cite{pereira2016}.

The experimental determination of graphene's thermal conductivity exhibits significant variance depending on the measurement technique. The reported values in the literature range from 1500 to 5000 W/mK \cite{balandin2008,ghosh2008,faugeras2010,cai2010,lee2011}. In contrast, the thermal conductivity of graphene derived from the REBO force field used in this study was approximately 1200 W/mK, as previously discussed \cite{mu2014,barbarino2015}. Classical force field-based methods consistently underestimate experimental or theoretically derived values, such as those obtained via the Boltzmann transport equation. Considering a reference thermal conductivity value of 2000 W/mK for graphene, the deviation introduced by the REBO force field approximates 40\%. Extrapolating this linear deviation to Irida-G, the thermal conductivity corresponding to this ``true'' value would be approximately 215 W/mK. We compared the value obtained to validate this estimate using a recently proposed method outlined in reference \cite{tromer2023}. This method provides rapid estimates of thermal conductivity based on semi-empirical calculations. The thermal conductivity at $T=300$ K is derived from an approximate expression:
\begin{equation}
\kappa_{\infty}=\frac{c~\bar{\omega}~E_\text{VIB}}{L~\delta T},
\label{kappa_fit}
\end{equation}
where $c$ is the speed of light in m/s, $\bar{\omega}$ is the average frequency, $E_\text{VIB}$ is the vibration activation energy obtained from the Arrhenius plot, $L$ is the lattice parameter of the minimum unit cell, and $\delta T$ is a free parameter that is equal to $3$ K for Irida-G, as discussed in reference \cite{tromer2023}. 

To apply the methodology outlined in reference \cite{tromer2023}, it is crucial to determine the values of the average $\bar{\omega}$ and the constant pressure heat capacity $C_{P,\text{VIB}}$ based on Irida's minimal unit cell. Employing the MOPAC2016 software, we identified the average $\bar{\omega}$ value, excluding negative frequencies as discussed in \cite{tromer2022}, as $1316.3$ cm$^{-1}$. By using the average lattice vectors $L=(l_x+l_y)/2.0=6.34$ \r{A} as the lattice parameter and substituting these parameters into Eq. (\ref{kappa_fit}), we estimated the thermal conductivity for Irida-G to be approximately $226.9$ W/mK. This value closely aligns with the corrected estimate for Irida-G, which was determined to be $215$ W/mK, using graphene's thermal conductivity as a reference.

\begin{figure}[t]
\centering
\includegraphics[width=\linewidth]{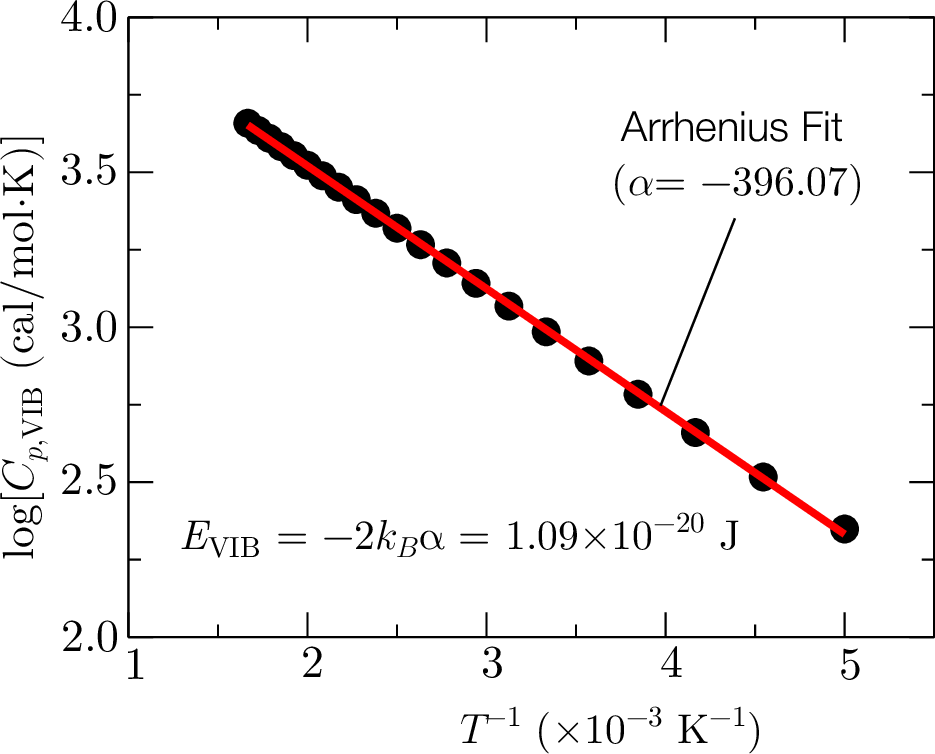}
\caption{Arrhenius plot for Irida-G using the method described in reference \cite{tromer2022}, where $\kappa_B$ represents the Boltzmann constant in Joules. The linear fit yields the slope $\alpha$, which determines the $E_{VIB}$ value.}
\label{fig:fit}
\end{figure}

It is worthwhile to stress that the thermal conductivity values reported for Irida-G in this study are comparable to those of other known 2D structures such as h-BN (220-550 W/mK) \cite{tabarraei2015thermal,yuan2019}, PHA-graphene (220-280 W/mK) \cite{pereira2016}, pentagraphene (167 W/mK) \cite{xu2015}, borophene-$\beta$ (90 W/mK) \cite{he2020}, and the biphenylene network (210-240 W/mK) \cite{ying2022}. Moreover, Irida-G exhibits a thermal conductivity considerably higher than that of the fullerene Network (4 W/mK) \cite{mortazavi2023}, WS$_2$ (32 W/mK) \cite{peimyoo2015}, MoS$_2$ (35 W/mK) \cite{yan2014}, silicene (9 W/mK) and germanene (2 W/mK) \cite{kuang2016}, phosphorene (14-30 W/mK) \cite{qin2015}, PtS$_2$ (85 W/mK) \cite{yin2021}, and others \cite{tromer2023}.

\subsection{Vibrational Spectrum}

To gain deeper insight into the mechanism behind the reduction in thermal conductivity of Irida-G compared to pristine graphene, we calculated the vibrational density of states (VDOS). The VDOS was derived through post-processing 100 ps trajectories with atomic velocities recorded every 5 fs. It was computed by taking the Fourier transform of the velocity autocorrelation function (VACF) as follows:
\begin{equation}
    \text{VDOS}(\omega)=\int_0^\infty \frac{\langle \textbf{v}(t)\cdot \textbf{v}(0) \rangle}{\langle \textbf{v}(0)\cdot \textbf{v}(0) \rangle} \exp{(-i\omega t)} dt
\end{equation}
\noindent where $\textbf{v}$ is the atomic velocity, $\omega$ is the angular frequency, and $\langle \textbf{v}(t) \cdot \textbf{v}(0) \rangle$ is the velocity autocorrelation function normalized such that $\text{VACF}(t=0)=1$.

\begin{figure}[t!]
\centering
\includegraphics[width=\linewidth]{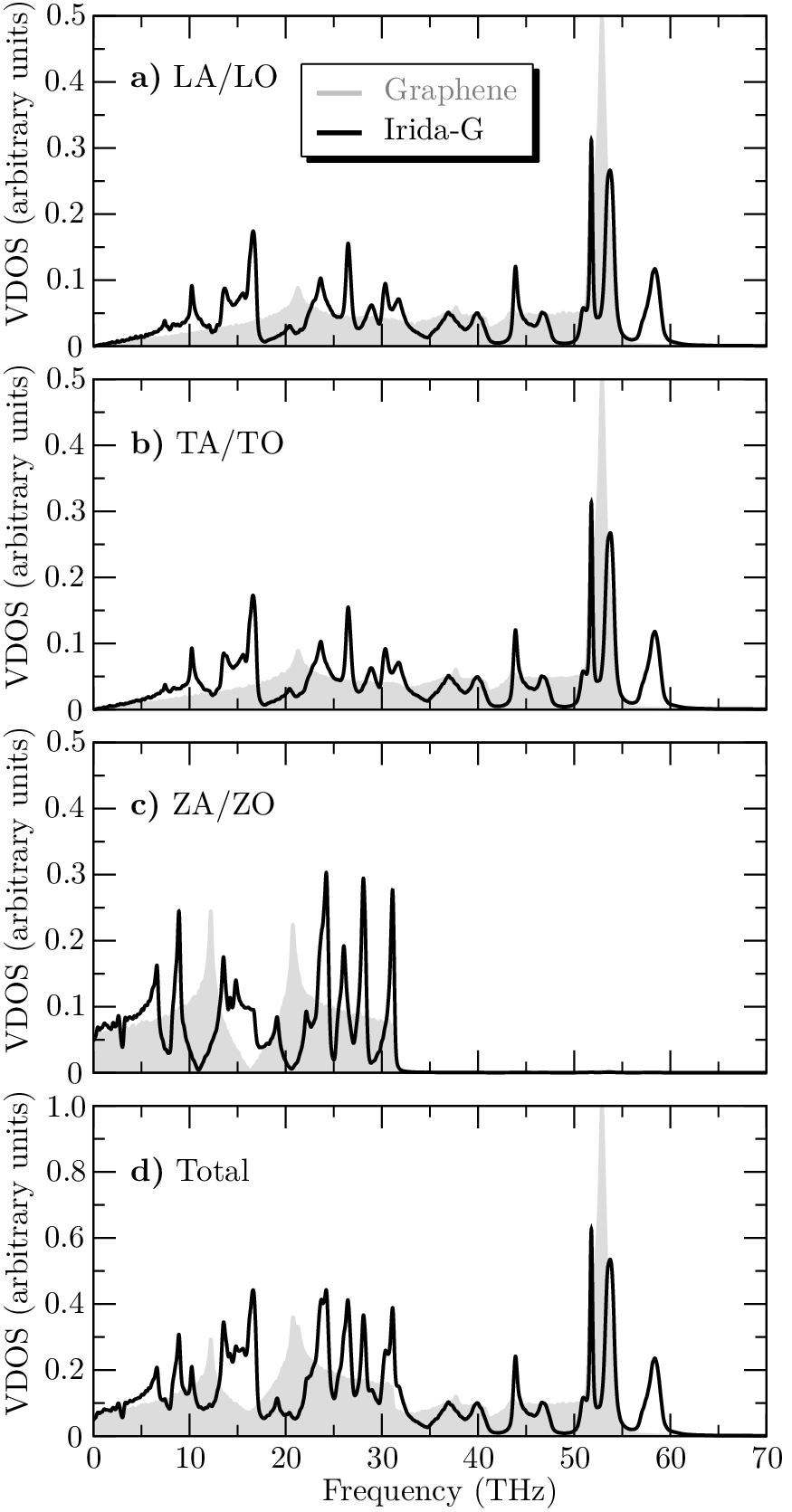}
\caption{Calculated vibrational density of states (VDOS) for pristine graphene (gray baseline) and Irida-G (black line). Panels (a), (b), and (c) illustrate the contributions from longitudinal (LA/LO), transverse (TA/TO), and flexural (ZA/ZO) modes, respectively. Panel (d) shows the total VDOS.}
\label{fig:vdos}
\end{figure}

Figure \ref{fig:vdos} clearly illustrates the different vibrational characteristics of the carbon atoms in Irida-G (black line) compared to graphene (gray baseline). Panel (a) shows the in-plane contributions of the longitudinal acoustic (LA) and longitudinal optical (LO) modes, while panel (b) depicts the in-plane contributions of the transverse acoustic (TA) and transverse optical (TO) modes. Panel (c) presents the out-of-plane contributions of the flexural acoustic (ZA) and flexural optical (ZO) modes. Finally, panel (d) displays the total VDOS.

Flexural phonons make the most important contributions to the VDOS in the low-frequency regime (below 10 THz), likely serving as the primary heat carriers in Irida-G. With their longer wavelengths, low-frequency phonons typically play a crucial role in heat transport in nanomaterials. These phonons undergo fewer scatterings and travel longer distances within the material, carrying significant energy and contributing more effectively to thermal transport. Moreover, the importance of flexural polarization (ZA) phonons, due to out-of-plane oscillations, in thermal transport has been well-documented for graphene \cite{lindsay2010_flexural,seol2010,pereira2013,barbarino2015}, and other 2D materials \cite{taheri2021}.

\subsection{Phonon Group Velocity}

\begin{figure}[t]
\begin{center}
\includegraphics[width=\linewidth]{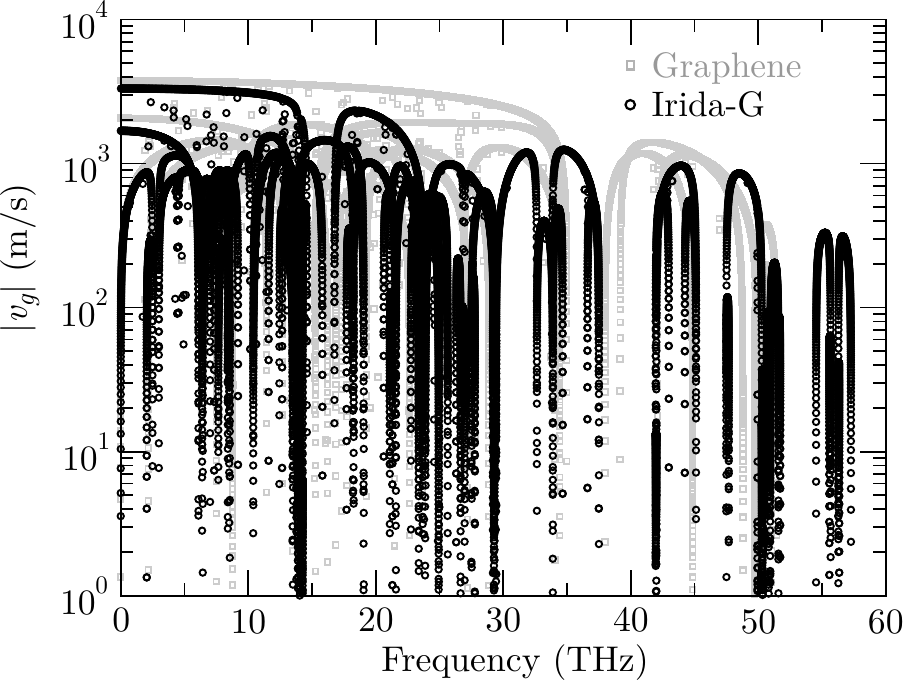}
\caption{Phonon group velocities as a function of frequency calculated from phonon dispersion relations using harmonic approximation.}
\label{fig:irida-gvel}
\end{center}
\end{figure}

To obtain a deeper understanding of the thermal transport characteristics of Irida-G, we also computed the phonon group velocities from the phonon dispersion relations within the harmonic approximation. The phonon group velocities are determined by:
\begin{equation}
    v_g (\omega) = \frac{d\omega}{dq}
\end{equation}
\noindent where $\omega$ represents the frequency of a given mode and $q$ denotes the wave vector. Figure \ref{fig:irida-gvel} illustrates the absolute values of the phonon group velocities as a function of frequency for Irida-G (black circles) and graphene (gray squares), both with the same number of atoms ($48$) and similar dimensions. Notably, data points between $\sim2\times10^3$ and $3\times10^3$ m/s, ranging up to 15 THz for Irida-G and up to 35 THz for graphene, correspond to the acoustic modes, which are the primary contributors to thermal transport, as previously discussed. The data indicate that the phonon group velocities of Irida-G are lower than those of graphene. The average group velocities are approximately 410 m/s and 780 m/s for Irida-G and graphene, respectively. Since lattice thermal conductivity is proportional to the square of the group velocities \cite{ziman2001}, this highlights the significant difference in thermal conductivity between these two materials. These results are consistent with the observed lower thermal conductivity of Irida-G relative to graphene.

\section{Conclusions}

In this study, we carried out molecular dynamics simulations to investigate the thermal transport properties of a newly proposed two-dimensional carbon nanomaterial, the so-called Irida-G. Our investigation focused on elucidating how the system's structural characteristics influence its thermal conductivity compared to graphene.

Initially, we verified the stability of Irida-G's crystal structure by calculating its phonon dispersion using the second-generation REBO potential. The absence of phonon modes with negative frequencies indicates the structural stability of the system, confirming that the chosen potential accurately describes the Irida-G structural features. Using the RNEMD method to create a temperature gradient within the system, our simulations unveiled a significant finding. Irida-G, at room temperature, exhibits an intrinsic thermal conductivity of approximately 130 W/mK. This value, notably lower than that of graphene, is primarily due to increased phonon scattering induced by the porous structure of Irida-G. Based on estimates derived from semi-empirical methods, we obtained a correction to the thermal conductivity value of the system, around 215 W/mK.

Furthermore, we analyzed the size dependence of thermal conductivity, revealing three distinct heat transport regimes: ballistic, diffusive, and a ballistic-diffusive transition regime. These regimes elucidate how the phonon mean free path influences heat transfer within Irida-G. The vibrational characteristics of Irida-G were obtained by calculating the VDOS and phonon group velocities. Our analysis revealed that flexural phonons, especially those with low frequencies, play a significant role in thermal transport within Irida-G. The lower phonon group velocities of the material compared to graphene further corroborate its reduced thermal conductivity.

\section*{Author contributions}
I.M.F.: Conceptualization, Methodology, Software, Validation, Formal analysis, Investigation, Data Curation, Writing - Original Draft, Visualization.
R.M.T.: Conceptualization, Methodology, Software, Validation, Formal analysis, Investigation, Data Curation, Writing - Original Draft.
L.D.M.: Formal analysis, Investigation, Resources, Funding acquisition, Writing - Review \& Editing.
D.S.G.: Formal analysis, Investigation, Writing - Review \& Editing, Funding acquisition.
L.A.R.J.: Formal analysis, Investigation, Writing - Review \& Editing, Funding acquisition.
M.L.P.J.: Conceptualization, Methodology, Formal analysis, Investigation, Resources, Writing - Review \& Editing, Supervision, Project administration, Funding acquisition.
All authors reviewed the manuscript. 

\section*{Conflicts of interest}
There are no conflicts to declare.

\section*{Data availability}
The data that support the findings of this study are available from the corresponding author, M.L.P.J., upon reasonable request

\section*{Acknowledgements}

This work received partial support from Brazilian agencies CAPES, CNPq, and FAPDF.
I.M.F. thanks the National Council for Scientific and Technological Development (CNPq) for grant no. 153604/2024-7.
R.M.T. acknowledges the support of the MackGraphe.
L.D.M. acknowledges the support of the High Performance Computing Center at UFRN (NPAD/UFRN).
D. S. G. acknowledges the Center for Computing in Engineering and Sciences at Unicamp for financial support through the FAPESP/CEPID Grant \#2013/08293-7.
L.A.R.J. acknowledges the financial support from FAP-DF 00193.00001808/2022-71 and 00193-00001857/2023-95 grant and FAPDF-PRONEM grant 00193.00001247/2021-20, and CNPq grant 350176/2022-1.
M.L.P.J. acknowledges the financial support of the FAP-DF grant 00193-00001807/2023-16. Thanks also to CENAPAD-SP (National High-Performance Center in São Paulo, State University of Campinas -- UNICAMP, project: proj960) and NACAD (High-Performance Computing Center, Lobo Carneiro Supercomputer, Federal University of Rio de Janeiro -- UFRJ, project: a22002) for the computational support provided. 
%

%%%END OF MAIN TEXT%%%


\balance


%%%REFERENCES%%%
\bibliography{main} %You need to replace "rsc" on this line with the name of your .bib file
\bibliographystyle{rsc} %the RSC's .bst file
\end{document}